\def \SAIT #1 #2 {{\em Mem.\ Soc.\ Astron.\ It.\/} {\bf #1}, #2}
\def \MESS #1 #2 {{\em The Messenger\/} {\bf #1}, #2}
\def \ASTRNACH #1 #2 {{\em Astron. Nach.\/} {\bf #1}, #2}
\def \AAP #1 #2 {{\em Astron. Astrophys.\/} {\bf #1}, #2}
\def \AAL #1 #2 {{\em Astron. Astrophys. Lett.\/} {\bf #1}, L#2}
\def \AAR #1 #2 {{\em Astron. Astrophys. Rev.\/} {\bf #1}, #2}
\def \AAS #1 #2 {{\em Astron. Astrophys. Suppl. Ser.\/} {\bf #1}, #2}
\def \AJ #1 #2 {{\em Astron. J.\/} {\bf #1}, #2}
\def \ANNREV #1 #2 {{\em Ann. Rev. Astron. Astrophys.\/} {\bf #1}, #2}
\def \APJ #1 #2 {{\em Astrophys. J.\/} {\bf #1}, #2}
\def \APJL #1 #2 {{\em Astrophys. J. Lett.\/} {\bf #1}, L#2}
\def \APJS #1 #2 {{\em Astrophys. J. Suppl.\/} {\bf #1}, #2}
\def \APSS #1 #2 {{\em Astrophys. Space Sci.\/} {\bf #1}, #2}
\def \ASR #1 #2 {{\em Adv. Space Res.\/} {\bf #1}, #2}
\def \BAIC #1 #2 {{\em Bull. Astron. Inst. Czechosl.\/} {\bf #1}, #2}
\def \JSQRT #1 #2 {{\em J. Quant. Spectrosc. Radiat. Transfer\/} {\bf #1}, #2}
\def \MN #1 #2 {{\em Mon. Not. R. Astr. Soc.\/} {\bf #1}, #2}
\def \MEM #1 #2 {{\em Mem. R. Astr. Soc.\/} {\bf #1}, #2}
\def \PLR #1 #2 {{\em Phys. Lett. Rev.\/} {\bf #1}, #2}
\def \PASJ #1 #2 {{\em Publ. Astron. Soc. Japan\/} {\bf #1}, #2}
\def \PASP #1 #2 {{\em Publ. Astr. Soc. Pacific\/} {\bf #1}, #2}
\def \NAT #1 #2 {{\em Nature\/} {\bf #1}, #2}

\documentstyle{memsait}
\input epsf.sty
\begin{opening}
\title{SIMULATING INHOMOGENEOUS REIONIZATION} 
\author{MICHAEL L. NORMAN$^{1,2}$, PASCALIS PASCHOS$^{1,2}$, TOM ABEL$^{1,3}$}
\institute{$^1$Laboratory for Computational Astrophysics, NCSA, Urbana, IL, USA\\
$^2$Astronomy Department, University of Illinois, Urbana, IL 61801, USA\\
$^3$Max-Planck-Institut f\"{u}r Astrophysik, Garching, Germany}
\date{} 
\end{opening}

\begin{document}

\oddpagefooter{}{}{} 
\evenpagefooter{}{}{} 
\ 
\bigskip

\begin{abstract}
We describe an approach for incorporating radiative transfer into 3D
hydrodynamic cosmological simulations. The method, while approximate,
allows for a self-consistent treatment of self-shielding and
shadowing, diffuse and point sources of radiation, and frequency
dependent transfer. Applications include photodissociation,
photoheating, and photoionization of the IGM.
\end{abstract}

\section{Introduction}
According to current thinking, the epoch of hydrogen reionization begins
with the formation of the first massive stars in subgalactic objects
at high redshifts $z \sim 15$ (e.g., Couchman \& Rees 1986; 
Gnedin \& Ostriker 1997; Haiman \& Loeb 1997) and is essentially 
complete by $z \sim 5$ as required by the Gunn-Peterson test. 
Due to its higher ionization threshold and lower photoionization
rate, helium reionization may be delayed to $z \sim 3$ or less
(Meiksin \& Madau 1993; Reimers et al. 1997). The essential reason
for this delay is that it is not until
$z \sim 4$ that the spectrum of the UV background has
hardened sufficiently due the rising quasar population. 

Reionization has been studied in a spatially averaged 
fashion by many authors (e.g, Shapiro \& Giroux 1987; Miralda-Escud\'{e} \&
Ostriker 1990, 1992; Meiksin \& Madau 1993; Giroux \&
Shapiro 1996; Haiman (these proccedings,
and references therein.) These studies reinforce Shapiro \&
Giroux's conclusion that quasars are insufficient
to reionize the universe at $z > 3.5$ as required by the Gunn-Peterson
measurement. (For a discussion of the uncertainties involved, see
Meiksin \& Madau 1993.) Barring exotica, such as decaying
neutrinos, that leaves only stellar sources of reionization at higher
redshifts. These stars will inevitably form in highly localized 
regions of space (i.e., galaxies.) Thus, cosmic reionization is 
intrinsically highly inhomogeneous. 

In this short contribution, we outline our intended approach to numerically 
simulate {\it inhomogeneous reionization} in three spatial dimensions. 
We seek efficient numerical methods for evolving the ionizing radiation
field in 3D which we may couple to hydrodynamic models of structure
formation (e.g., Abel, Bryan \& Norman, these proceedings.)
We warn the reader that at this time, we have not proven that our method 
will be sufficiently accurate and robust to meet our
goal, nor do we have any measurements of the computational cost.

\section{Description of the Problem}

Immediately after the first photoionizing sources turn on, one
has a collection of isolated ionization zones (essentially
HII regions) growing in an inhomogeneous, expanding universe. 
Their number and evolution will depend on the source population,
ionizing fluxes, and ambient conditions. Until the HII regions
begin to interact (percolation), they can be solved as isolated
cases. The homogeneous case has been solved by Shapiro (1986) for the case 
of quasar reionization.  With no inhomogeneities
to spoil spherical symmetry or emit appreciable ionizing recombination
radiation, one can reduce the problem to
an ODE for the radius of the I-front versus redshift.
One avoids solving the radiative transfer equation by simply
attenuating the ionizing flux by the $1/r^2$ geometrical factor and
volumetric losses due to recombination to atomic levels $n > 1$.
The peculiar velocity of the I-front is set by 
balancing, in the rest frame of the I-front, outgoing ionizing photons
and incoming neutrals. With these approximations, Shapiro found that for
typical quasar luminosities, the I-front always expands 
supersonically with respect to both neutral and ionized media.
Such I-fronts are known as weak R-type I-fronts (cf. Spitzer 1978). Thus
hydrodynamic motions are unimportant, justifying their neglect  
{\it a posteriori}. Secondly, he found that 
the I-fronts never reach their equilibrium Str\"{o}mgren radii in a Hubble
time, although they do overlap completely by $z = 0$ (but not by $z=5$)
for the observed 
number density of quasars. This is simply because the recombination time, 
which enters in the definition of the Str\"{o}mgren radius, is longer than 
the Hubble time, and consequently the Str\"{o}mgren radius is very large.

The evolution of an HII region in an inhomogeneous medium (expanding
or otherwise) in 3D is an unsolved problem---one that requires more
powerful numerical methods to solve in the general case. The principal
complication is that localized density enhancements will both absorb the
primary ionizing flux (shadowing), and isotropically emit 
ionizing radiation via recombinations to the ground state (diffuse radiation.)
As a consequence, the ionizing radiation field becomes inhomogeneous
and anisotropic. The presence of density enhancements (clouds) retard the
expansion of the I-front relative to the homogenous case due to
enhanced down-conversion of ionizing photons into non-ionizing Balmer
continuum photons. 
When the areal covering factor of opaque clouds approaches one, the I-front
will become starved of ionizing photons and stop expanding.  Well before
this limit is reached, hydrodynamics will become important for two reasons.
First, dense clouds inside the HII region will be photo-evaporated, 
leading to peculiar velocities of order the thermal sound speed 
$\sim 30 ~km/s$ (see Shapiro, these proceedings.)
Secondly, when the I-front expansion rate falls below the
sound speed in the HII region, it makes a transition to a weak D-type
front (cf. Yorke 1986). When this occurs, the pressure difference 
between the ionized gas and the neutral gas drives a shock wave into the 
ambient medium. The subsequent expansion of the HII region is due to a 
combination of hydrodynamic and radiative effects. A shell of material
is accumulated between the inner I-front and the outer shock front, which
separate from one another. 

The complexities of the percolation phase depend on whether hydrodynamic
effects are imporant or not. That in turn depends on the presence of
density inhomogeneities and whether the isolated HII regions become
weak D-type before or after overlap. 
Shapiro \& Giroux (1987) modeled the effects of density inhomogeneities
on quasar-driven I-fronts phenomenologically by adding a clumping
factor to the Shapiro (1986) formalism. Physically, these clumps
correspond to the $Ly \alpha$ forest and Lyman limit systems. 
Meiksin \& Madau (1993) 
calculate that the clouds may absorb as much as $50\%$ of the UV
radiation from point sources. Consequently, the clouds are a substantial
diffuse source of ionizing radiation (Haardt \& Madau 1996.)
Shapiro \& Giroux (1987) showed that although the enhanced recombinations
increase the UV background deficit relative to observed QSO source population,
I-fronts remain supersonic (globally) until full overlap is achieved. 

However, if quasars fade, or if reionization is caused by far more numerous, 
less luminous stellar UV sources, hydrodynamics will become important as 
I-fronts
become weak D-type as they approach their Str\"{o}mgren radii.  
We desire numerical methods which can treat both kinds of circumstances.

\section{Basic Approach}

The basic elements of our method can be described very simply.
In the next section we provide equations. We decompose the 
radiation field into point source and diffuse components: 
$I_{\nu} \equiv I^{pts}_{\nu} +
I^{diff}_{\nu}$. The point source radiation field is attenuated
along radial rays from each point source in the volume. Every
cell is crossed by at least one ray. The 
number of photoionizations in frequency interval $\nu, \nu + d\nu$ 
in each ray segment inside a cell is simply related to the
decrease in $I^{pts}_{\nu}$ along that segment.
The total number of photoionizations in the cell is the sum over
ray segments within the cell. The diffuse radiation field is computed
by solving the angle-integrated moment equations
with appropriate Eddington factor closure (Stone, Mihalas \& Norman
1992). In the limit of small (compared to the horizon) volumes,
the zeroth and first moment
equations can be simplified and combined into a single nonlinear
elliptic equation for  $J^{diff}_{\nu} \equiv \frac{1}{4\pi}
\int I^{diff}_{\nu} d\Omega $. 
This elliptic equation is discretized on a 3D cartesian mesh
and may be solved using a variety of techniques, including
multigrid relaxation and/or iterative sparse-banded matrix
methods.

\section{Formalism}

\subsection{Equation of Cosmological Radiative Transfer}     

The equation of cosmological radiative transfer in comoving coordinates
(cosmological, not fluid) is  (Paschos, Mihalas \& Norman 1998):

\begin{equation}
\frac{1}{c} \frac{\partial I_{\nu}}{\partial t} + \frac{\hat{n} \cdot
\nabla I_{\nu}}{\bar{a}} - \frac{H(t)}{c} (\nu \frac{\partial I_{\nu}}
{\partial \nu} - 3 I_{\nu}) = \eta_{\nu} - \chi_{\nu} I_{\nu}
\end{equation}
where $I_{\nu} \equiv I(t, \vec{x}, \vec{\Omega}, \nu)$ is the monochromatic
specific intensity of the radiation field, $\hat{n}$ is a unit vector
along the direction of propagation of the ray; $H(t) \equiv \dot{a}/a$ is the 
(time-dependent) Hubble constant, and $\bar{a} \equiv \frac{1+z_{em}}{1+z}$ is the 
ratio of cosmic scale factors between photon emission at frequency
$\nu$ and the present time t.
The remaining variables have their traditional meanings (e.g, Mihalas 1978.)
Equation (1) will be recognized as the standard equation of radiative
transfer with two modifications: the denominator $\bar{a}$ in the 
second term, which accounts for the changes in path length along
the ray due to cosmic expansion, and the third term, which 
accounts for cosmological redshift and dilution.

In principle, one could solve equation (1) directly for the intensity
at every point given $\eta$ and $\chi$. However the high dimensionality
of the problem (three positions, two angles, one frequency and time) 
not to mention the high spatial and angular resolution needed 
in cosmological simulations make this approach impractical for dynamic 
computations. Therefore
we proceed through a sequence of well-motivated approximations which
reduce the complexity to a tractable level.

\subsection{Local Approximation}

We begin by eliminating the cosmological terms and factors. That we can
do this can be understood on simple physical grounds. Before the universe
is reionized, it is opaque to H and He Lyman continuum photons. Consequently,
ionizing sources are local to scales of interest, and not at cosmological 
distances. If our simulation box is of side length L and $\lambda_p $ is 
the photon mean free path, then by construction $\lambda_p \ll L$.
The ratio of the third to the second terms in
equation (1) is $HL\bar{a}/c \ll 1$, and hence the third term can
safely be ignored. Now, let us consider the factor $\bar{a}$ in
equation (1). For a photon which is emitted at time t on one side of the 
box and absorbed on the other side at time $t + L/c$, 
$ \bar{a} = (\frac{t+l/c}{t})^\eta \sim 1 + \eta L/ct = 1 + \eta L/L_H$,
where $\eta$ is the logarithmic expansion rate of the universe (2/3 for
$\Omega_o = 1$) and $L_H$ is the Hubble horizon scale. 
For $L \ll L_H, \bar{a} \doteq 1$, and $\nu_{em} \doteq \nu$.
In practice, our dynamical timesteps are much longer than a photon
crossing time. However, even in this case accuracy limits our dynamical 
timesteps such that $\Delta a
/a \ll 1$, and hence $\bar{a} \doteq 1$ in any given timestep. 
Therefore, setting $\bar{a} \equiv 1$, equation (1) 
reduces to its standard, non-cosmological form:

\begin{equation}
\frac{1}{c} \frac{\partial I_{\nu}}{\partial t} + \hat{n} \cdot
\nabla I_{\nu} = \eta_{\nu} - \chi_{\nu} I_{\nu}
\end{equation}
\noindent
where now $\nu$ is the instantaneous, comoving frequency.

Now consider the case where $\lambda_p \gg L$, i.e., 
the simulation volume is optically thin to ionizing radiation
as it would be after reionization. In that case ionizing sources are
are either inside the box (local) or outside the box (nonlocal), or both.
Local sources are treated as in the case above.
Nonlocal sources sufficiently far from the box contribute to a nearly 
isotropic, homogeneous UV
background (metagalactic flux). In this limit, equation (1) can be
solved in an angle and spatially averaged fashion, and the cosmological
term is not ignorable. This computation 
has been done by Haardt \& Madau (1996)
including emission from the observed quasar population, and absorption
and re-emission by the $Ly \alpha$ forest. The result of this calculation
is $J^*_{\nu}(z)$---the mean metagalactic intensity as a function of
redshift. If the material in the simulation volume is optically
thin everywhere, then the local radiation field is $J^*_{\nu}(z)$.  
Current simulations of the $Ly \alpha$ forest (e.g., Zhang et al. 1997, 
1998) employ this approximation. If, however, there exist opague
regions within the simulation volume---for example, high column density
Lyman alpha clouds---then $J_{\nu}$ no longer equals $J^*_{\nu}(z)$
locally, but must be computed using some approximation to
equation (1) using $J^*_{\nu}(z)$ as a boundary
condition. In this case, the cosmological terms are accounted for in
the boundary conditions, and for box sizes much smaller than the
horizon, ignorable within the box.  

\subsection{Angular Moments} 

While the radiation field due to local point sources is highly anisotropic, 
the diffuse radiation field should be more nearly isotropic since 
recombination radiation is emitted isotropically and absorbed by density 
enhancements which are well resolved on our computational grid. Thus we 
expect the angular structure in $I^{diff}_\nu$ to be well described by 
its angular moments, the first three of which are defined as follows: 
$J_\nu \equiv \frac{1}{4\pi}\oint d\Omega I_\nu;
K^i_\nu \equiv \frac{1}{4\pi}\oint d\Omega n^i I_\nu; 
K^{ij}_\nu \equiv \frac{1}{4\pi} \oint d\Omega n^i n^j I_\nu$.  
The radiation energy density, flux, and pressure tensor are related to these 
moments via the simple relations $E_\nu = \frac{4\pi}{c}J_\nu, 
F^i_\nu = \frac{4\pi}{c}H^i_\nu,$ and $P^{ij}_\nu = \frac{4\pi}{c}K^{ij}_\nu.$ 

Now, it is advantageous to work in a frame which is comoving with the fluid 
because in this frame the emission and absorption coefficients are isotropic 
(Mihalas \& Mihalas 1984; hereafter MM).  
Denoting {\em comoving frame} quantities with a subscript ``o",   
The first two moment equations of the radiation field
are obtained by taking angular moments of equation (2), which yields
temporarily suppressing the $\nu$ subscript:

\begin{equation}
\rho \frac{D}{Dt} (\frac{E_o}{\rho}) + \nabla_i F^i_o + P^{ij}_o \nabla_i v_j = 
4\pi \eta_o -c \chi_o E_o,
\end{equation}
\begin{equation}
\frac{\rho}{c^2} \frac{D}{Dt} (\frac{F^i_o}{\rho}) + \nabla_j P^{ji}_o + 
\frac{1}{c^2} F^j_o \nabla_j v^i = -\frac{1}{c}\chi_o F^i_o,
\end{equation}

\noindent
where $\rho$ is the fluid density and $\frac{D}{Dt} \equiv \frac{\partial}{\partial t} + 
\vec{v} \cdot \vec{\nabla}$ is the convective derivative. 
Equations (3) and (4) are MM equations (95.87) and (95.88) dropping acceleration terms.

\subsection{Quasi-Static Approximation}

Equations (3) and (4) can be simplified further when we realize we are interested
in phenomena occuring on the fluid flow timescale, not the radiation flow timescale.
This {\em quasi-static approximation} amounts to throwing away terms which are
always $O(\frac{v}{c})$ or higher. This must be 
done with care, and we refer the reader
to MM for a thorough analysis. The result, for {\em continuum
radiation} is:

\begin{equation}
\nabla_i F^i_\nu = \epsilon_\nu -c k_\nu(t)E_\nu
\end{equation}
\begin{equation}
\nabla_j P^{ji}_\nu = -\frac{k_\nu(t)}{c} F^i_\nu
\end{equation}

\noindent
where now it is understood all quantities are measure in the comoving (fluid) frame, 
$\epsilon_\nu = 4\pi\eta_\nu$, and $k_\nu$ is the absorption coefficient
(we have ignored scattering.) Equations (5) and (6) are called quasi-static
because the only timescale which enters is through the
material opacity $k_\nu(t)$, which evolves  
on a photoionization timescale.

\subsection{Closure Schemes}

Equations (5) and (6) are two equations in three unknowns: E, F and P.
We will experiment with various approaches to closing the hierarchy of
moment equations. The most general and accurate approach is the 
{\em Eddington factor closure}: $P^{ij}_\nu = f^{ij}_\nu E_\nu$, where
$f^{ij}_\nu$ is the tensor Eddington factor (e.g., Stone, Mihalas \&
Norman 1992). $f_\nu^{ij}$ contains all the angular information of the
radiation field. If one knew $I_\nu(\Omega, \vec{x})$, say from solving equation
(2), then $f^{ij}$ can be computed via angular quadratures at every 
point $\vec{x}$. However, the goal is to avoid solving the angle-dependent 
transfer equation. One approach, which we call NEWS, is to 
compute $I_\nu$ only along rays parallel to grid lines, as well
as along the principal diagonals. Since we use uniform cartesian meshes
in cosmology, we simply evaluate the formal solution to eq. (2) on long
characteristics. A second, less accurate approach, is to compute $f^{ij}$
using geometric information about the location of principal emitters
and absorbers. This approach would have to be calibrated against more exact
methods.

Yet more approximate, but perhaps adequate for our needs, is the {\em diffusion
approximation} which states that $F^i_\nu = -D\nabla^iE_\nu$, where D depends
on the energy density and opacity through the relation: $D=\frac{c}{k_\nu}
\lambda(E_\nu)$. The quantity $\lambda$ is called the flux limiter, and for
optically thick media has a value of 1/3.
When radiation propagates through optically thin media, it streams rather than
diffuses. There are many functional forms for $\lambda$ which are taylored
to the problem under study; we mention three which are prominent in the
literature by Alme \& Wilson (1974), Minerbo (1978) and Levermore \& Pomraning
(1981). All depend on the quantity $R = \frac{|\nabla E_\nu|}{k_\nu E\nu}$,
which is used as a switch between the diffusion and free streaming limit. 
Whether any of these formulations will prove adequate for our application
remains to be seen.

Inserting the relation $F^i_\nu = -D\nabla^iE_\nu$ into equation (5), we
obtain
\begin{equation}
\nabla_i (\frac{\lambda}{k_\nu}\nabla^i E_\nu) = k_\nu (\Im_\nu -  E_\nu),
\end{equation}

\noindent
where $\Im_\nu = \frac{\epsilon_\nu}{c k_\nu} = \frac{4\pi}{c}S_\nu$, S: the  
source function. Since $\lambda = \lambda(E_\nu, \nabla E_\nu)$,  
equation (7) is a nonlinear, elliptic equation for $E_\nu$, where the 
quantities $\Im_\nu$ and $k_\nu$ are functions of space and time. Solution
requires the specification of boundary values on $E_\nu$ and its normal
derivative.

\subsection{Frequency Reduction}

Finally, we consider reducing the frequency complexity. For this, we
employ the {\em multigroup method} (cf. MM, Ch. 6), 
in which the frequency spectrum is 
divided into a number of {\em frequency groups}. Since, in the first
instance,  we are are only interested in the photoionization of primordial
gas, we need only consider three frequency groups above the
ionization edges for HI, HeI and HeII at energies $h\nu$ = 1, 1.809, and 4 Ryd, 
respectively. Defining the group average radiation energy density as:

\begin{equation}
E_g \equiv \int_{\nu_g}^{\nu_{g+1}} E_\nu d\nu / (\nu_{g+1} - \nu_g)
\end{equation}
\noindent
the multigroup diffusion equations to be solved are:
\begin{equation}
\nabla_i (\bar{D_g}\nabla^i E_g) = \epsilon_g - c \bar{k_\nu} E_g,
\end{equation}
\noindent
where $\bar{D_g} \equiv \frac{\int_{\nu_g}^{\nu_{g+1}} D_\nu \nabla E_\nu d\nu}
{\int_{\nu_g}^{\nu_{g+1}}\nabla E_\nu d\nu}$
and
$\bar{k_g} \equiv \frac{\int_{\nu_g}^{\nu_{g+1}} k_\nu E_\nu d\nu}
{\int_{\nu_g}^{\nu_{g+1}} E_\nu d\nu}$, the flux mean and absorption mean
opacities, respectively. Because we don't know $E_\nu$ and $\nabla E_\nu$ in
advance, these mean opacities must be approximated. 
The approximations, while not
unique, can be constructed to have various desirable properties, such as 
giving the exact energy or momentum absorbed within a group. 
We desire group means which conserve the total number of photoionizations 
within a group to a high degree of accuracy.
Thus, we write:
\begin{equation}
k_{ph} = \int_{\nu_L}^\infty d\nu \frac{c E_\nu}{h\nu}(\sigma_\nu^{HI}+
\sigma_\nu^{HeI}+\sigma_\nu^{HeII}) = \sum_{g=1}^3 \frac{c E_g}{h\nu_{g+1/2}}
<\sigma_\nu^{HI}+\sigma_\nu^{HeI}+\sigma_\nu^{HeII}>_g(\nu_{g+1} - \nu_g)
\end{equation}
\noindent
where $\nu_L$ is the Lyman limit, and $\sigma^x_\nu$ is the photoionization
cross section for species x. The angle average is defined formally as:
\begin{equation}
<Q>_g = \int_{\nu_g}^{\nu_{g+1}} d\nu W_g(\nu)Q(\nu)/(\nu_{g+1} - \nu_g),
\end{equation}
where the weighting function $W_g(\nu)$ depends on the assumed spectral
form (e.g., piecewise constant) of $E_\nu$ within each frequency group.

\section{Coupling to Chemistry and Hydrodynamics}

Here we briefly discuss the solution strategy for the coupled radiation,
matter, cosmological fluid dynamical system. A method for efficiently computing
nonequilibrium ionization, chemistry, and cooling coupled to mulitspecies
cosmological hydrodynamics is described in Anninos et al. (1997). 
Applications of this method to first structure formation and the
$Ly \alpha$ forest are found in Abel et al. (1998) and Zhang et al. (1998),
respectively. In the latter, photoionization of the IGM in the optically 
thin limit due to a metagalactic UV background is included. At the heart
of the method is an implicit scheme for solving the kinetic equations for
all the ionization states of H and He, as well as the reactants for $H_2$.
Inside the main loop of the hydrodynamic computation, we subscycle on
heating and cooling portion of the gas energy equation and the kinetic
equations. The timestep during subcyclying is chosen such that the
fractional abundances of species which dominate the cooling change by
no more than 10\%. Since all atomic and molecular cooling is proportional
to the electron density, we find it sufficient to limit our chemistry
timestep to $\Delta t_e = 0.1 n_e/\dot{n}_e$. 

Radiative transfer merely changes {\em when and where} material gets
photo- heated, ionized and dissociated. Thus, within the chemistry/cooling
subcycle loop we also call the methods described above for computing the 
point source and diffuse radiation field.
Since opacities change on the photoionization timescale which is reflected
in the electron fraction, we need not change our subcyle timestep
criterion to include radiative transfer.

\acknowledgements 
We thank the organizers F. Palla, D. Galli and E.  Corbelli for a most
enjoyable meeting in Florence. One of us (T.A.)  thanks D. Mihalas for
suggesting the NEWS scheme. The work was partially support by grants
NSF ASC-9318185 and NASA NAG5-3923.


\end{document}